\documentclass[aps,prb,twocolumn,groupedaddress,showpacs,amsmath,floatfix]{revtex4}
\usepackage{graphicx}   
\begin{document}

\title{Crossover from time-correlated single-electron tunneling to that of Cooper pairs}

\author{Jonas Bylander}\email[jonas.bylander@mc2.chalmers.se]{}
\author{Tim Duty}\altaffiliation{Present address: Australian Research Council Centre of Excellence for Quantum Computer
Technology, University of New South Wales, Sydney, NSW 2052,
Australia}\email[timd@phys.unsw.edu.au]{}
\author{G{\"o}ran Johansson}
\author{Per Delsing}
\affiliation{Microtechnology and Nanoscience, MC2, Chalmers
University of Technology, SE-412 96 G{\"o}teborg, Sweden}

\date{\today}

\begin{abstract} We have studied charge transport in a one-dimensional chain of
small Josephson junctions using a single-electron transistor. We
observe a crossover from time-correlated tunneling of single
electrons to that of Cooper pairs as a function of both magnetic
field and current. At relatively high magnetic field,
single-electron transport dominates and the tunneling frequency is
given by $f=I/e$, where $I$ is the current through the chain and $e$
is the electron's charge. As the magnetic field is lowered, the
frequency gradually shifts to $f=I/2e$ for $I\gtrsim200$\,fA,
indicating Cooper-pair transport. For the parameters of the measured
sample, we expect the Cooper-pair transport to be incoherent.
\end{abstract}


\maketitle

Charge transport in 1D and 2D arrays of small Josephson junctions
exhibits a wide range of physical
phenomena.\cite{Fazio-vdZant-PhysRep-01} In these systems there is a
competition between the Coulomb blockade, which tends to localize
charge, and the Josephson effect, which tends to delocalize it.
Depending on the parameters of the Josephson junctions in the array,
the transport can be described in terms of either vortices or
charges, which are dual entities in a superconducting system in the
sense that phase and charge are conjugate variables. For strong
Josephson coupling, $E_J$, the transport is better described in
terms of vortices. On the other hand, if the charging energy, $E_C$,
is larger, the system is better described in terms of charge
transport. This duality is not perfect since charge can be carried
by either Cooper pairs or electrons, whereas there is only one type
of vortex. Thus the competition between Cooper-pair tunneling and
single-electron tunneling is of particular interest.

We have previously demonstrated that single electrons can be counted
one by one as they tunnel through a 1D series-array of small
metallic islands connected by Josephson
junctions.\cite{Bylander-Nature-05} As one excess electron charges
an island in the array, it polarizes the neighboring islands and
forms a single-charge
``soliton."\cite{Likharev-IEEETransMag25-89,BenJacobMullenAmman-PLA-89}
Different solitons affect each other by Coulomb repulsion, and
therefore they form a 1D Wigner-like lattice that moves along the
array. This spatial separation enables a detector to resolve the
individual charges as they pass by. Moreover, their passage is time
correlated\cite{BenJacobGefen-PLA-85,Averin-JLTP62-86,Delsing-PRL63-1861-89}
with the frequency $f=I/e$, where $I$ is the current and $e$ the
electron's charge. As detector, we used a single-electron
transistor\cite{Likharev-IEEE-mag-87,FultonDolan-PRL-87} (SET)
connected to the end of the array.

\begin{figure}[b]
\vspace{-.7cm} \centering
 \includegraphics[height=0.45\columnwidth]{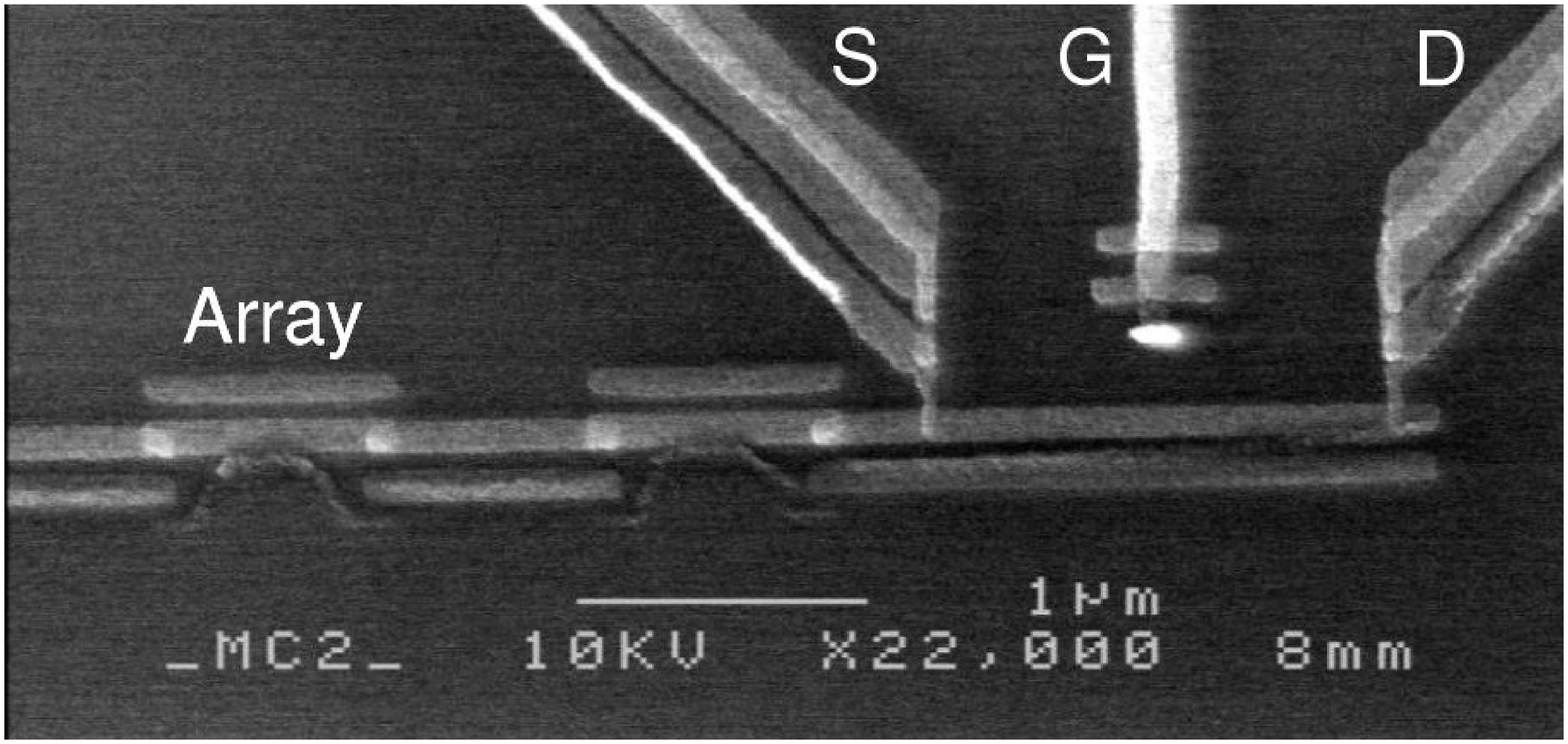}\vspace{-1.6cm}
 \includegraphics[height=0.55\columnwidth]{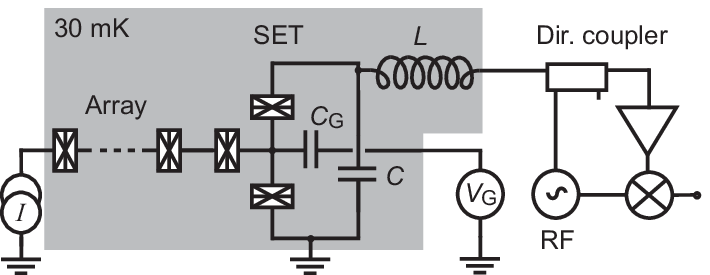}
 \caption{(a) Scanning electron micrograph of the sample. The last few islands of the 50-junction
 array are shown. The SET's source, drain and gate electrodes are labeled S, D and G, respectively.
 (b) Simplified circuit diagram.
 The charge entering the SET island modifies the dissipation of the $LC$ resonator,
 which is detected by RF reflectometry.
 }
\label{fig:SEMpicture}
\end{figure}

In Ref.~\onlinecite{Bylander-Nature-05} we only discussed the single
electron transport. However, since the array is superconducting, the
current can be carried either by electrons or by Cooper pairs.
In this paper, we report new results from measurements on the same
device (Fig.~\ref{fig:SEMpicture}), where we now study the
competition between single-electron tunneling and single Cooper-pair
tunneling. We demonstrate a crossover from single-electron transport
to Cooper-pair transport as a function of magnetic field and
current.

The studied array is in the strong charging limit, $E_C \gg E_J$,
and is consequently best described in terms of charge transport.
Apart from single electron tunneling, we can in principle have two
different kinds of Cooper-pair tunneling: coherent and incoherent.
The former is equivalent to Bloch
oscillations,\cite{Likharev-JLTP59-85,Kuzmin-PRL67p2890-91} where
the system adiabatically follows the lower energy band of each
junction without dissipation. The latter involves transitions to
excited states and exchange of energy with the
environment.\cite{Devoret-PRL-90,Averin-PhysB-90} In this particular
sample, $E_J$ is smaller than the thermal energy $k_B T$, and the
system can therefore easily be excited, leading to dissipation and
loss of coherence. Thus any Cooper-pair transport in our device
should be predominantly incoherent.

\begin{figure}[t]
\centering
 \includegraphics[height=1.2\columnwidth]{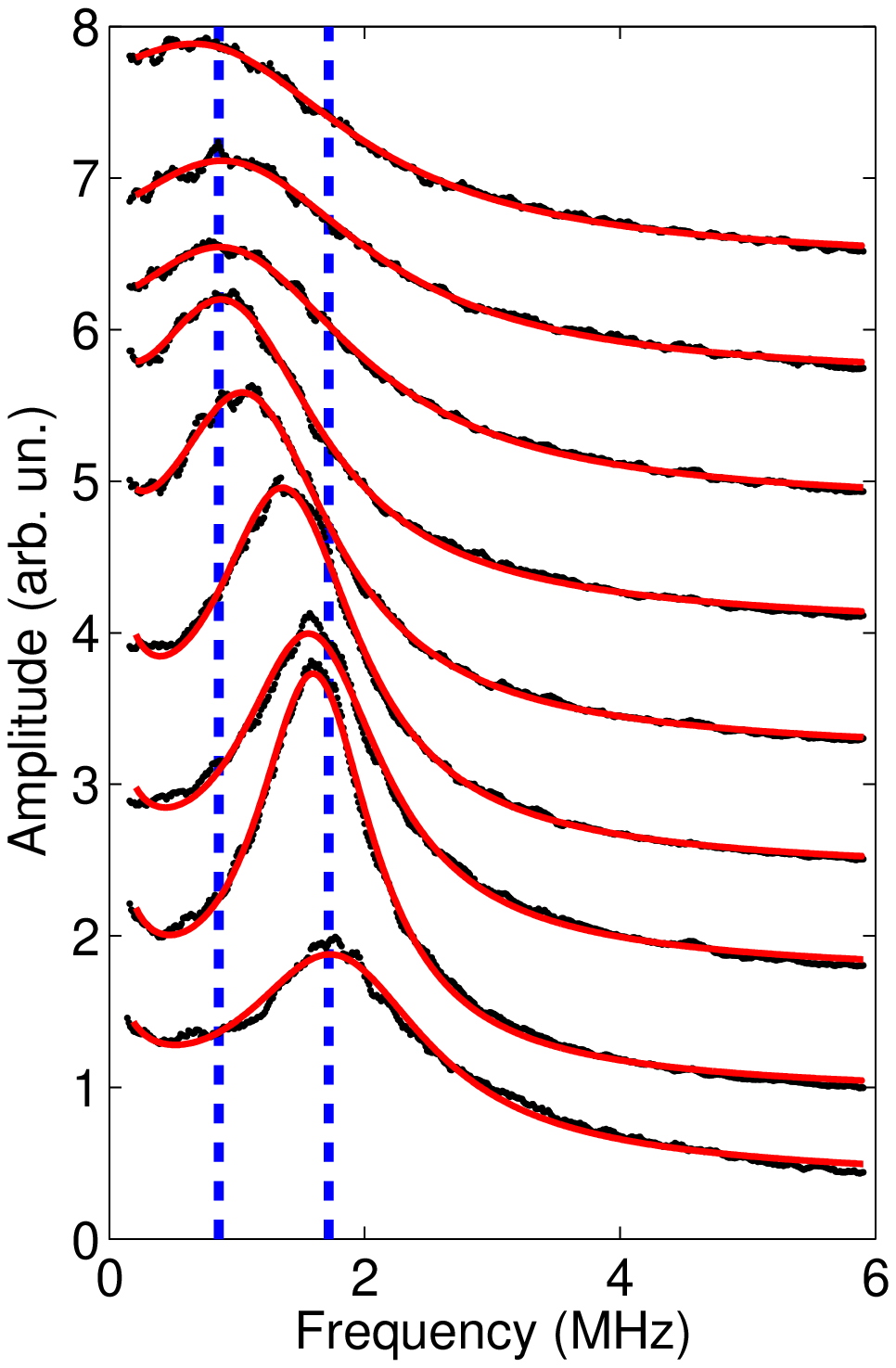}
 \includegraphics[height=.6\columnwidth]{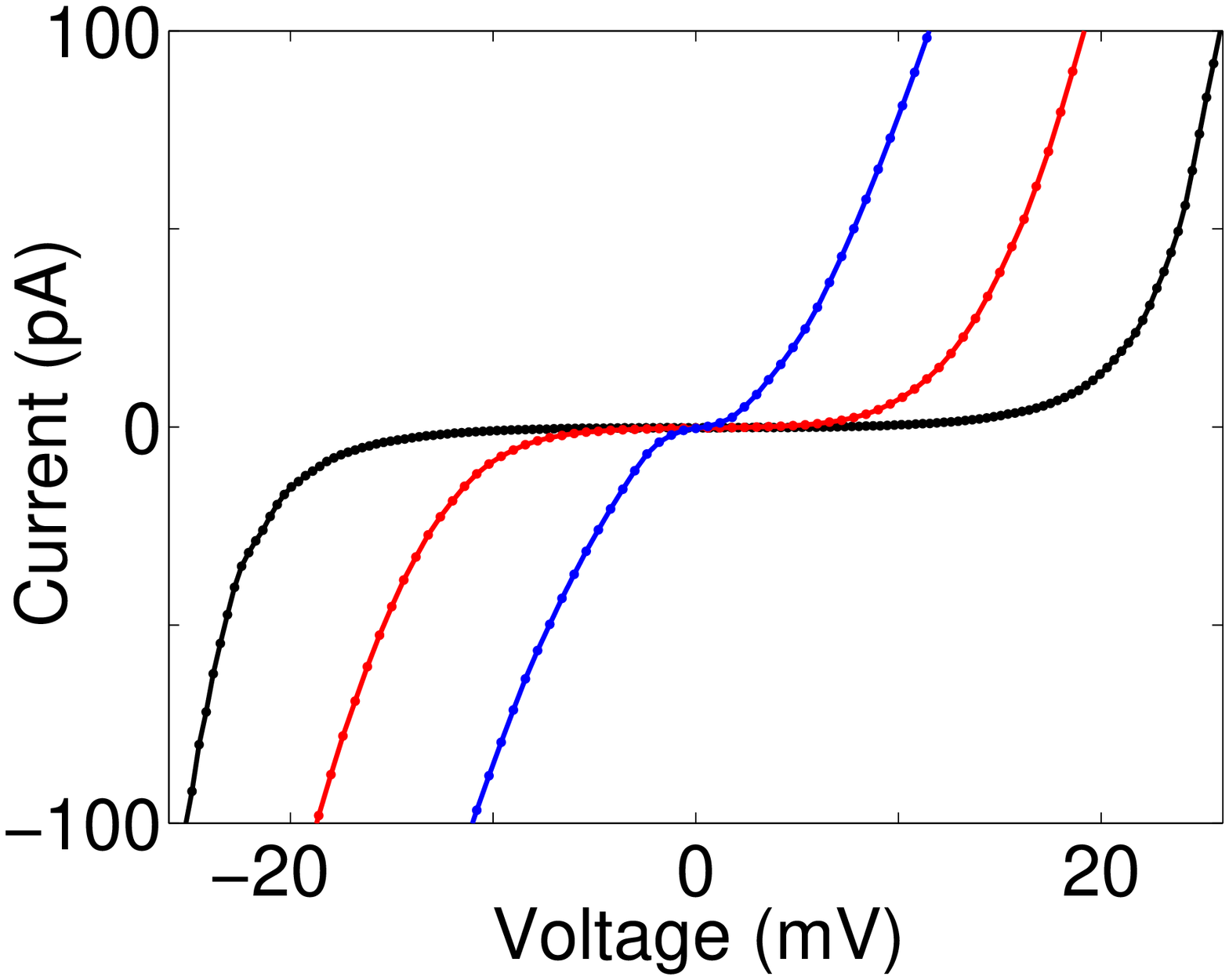}
 \caption{(color online) Upper graph: Power spectral density of the output signal from the
 RF-SET when the current through the array is maintained at 275 fA. The curves
 have been displaced vertically for clarity.
 Black lines are data starting at $B_{||}=500$\,mT (bottom) and
 continuing every 50\,mT to $B_{||}=100$\,mT (top). The solid (red) lines are fits
 to a Lorentzian plus a $\sim\!1/f$ background. The two dashed
 (blue) lines correspond to $f=I/e=1.72$\,MHz and $f=I/2e=0.86$\,MHz.
 Lower graph: Array $I-V$ curves at $B_{||}=0$ (black), $400$ (red), and $800$\,mT (blue).
 }
\label{fig:Peaks}
\end{figure}

We fabricated the sample using e-beam lithography and triple angle
evaporation of aluminum, allowing us to use different oxidation
parameters for the array and SET junctions. The average normal state
resistance for each of the $N=50$ array junctions was
$940\mathrm{\,k}\Omega$, which gives $E_J/k_B\approx$~10\,mK. The
charging energy per junction was $E_C/k_B=2.2$\,K, corresponding to
a junction capacitance of $C=0.42$\,fF. The capacitance to ground of
each island was $C_0\approx0.03$\,fF, giving a single-electron
soliton size $\Lambda\approx\sqrt{C/C_0}\approx4$, which is the
number of islands over which the array is polarized by a single
excess charge. The SET source--drain resistance was
$30\mathrm{\,k}\Omega$. A SEM picture of a sample is shown in
Fig.\,\ref{fig:SEMpicture}(a), and array current--voltage
characteristics in the lower graph of Fig.~\ref{fig:Peaks}.

We performed the measurements in a dilution refrigerator at
approximately 30\,mK. A magnetic field of up to 3\,T could be
applied parallel to the substrate. We determined the parallel
critical field for our sample from $I-V$ curves of the SET to be
$B_{||,c}\approx650$\,mT.

The SET was embedded in an $LC$ circuit and operated in the
radio-frequency mode (RF-SET);\cite{Schoelkopf-Science-98} the
resonance frequency was 358\,MHz and the bandwidth 10\,MHz. The
circuit's reflection coefficient depends sensitively on the charge
induced on the SET island. After amplification by cold and room
temperature amplifiers, the reflected signal was demodulated by
homodyne mixing and the baseband signal was then measured by a
spectrum analyzer, see Fig.\,\ref{fig:SEMpicture}(b). The charge
sensitivity is, in general, magnetic field dependent, but was
typically $20~\mu e/\sqrt{\mathrm{Hz}}$ in our measurement.

The array was biased using a Keithley 263 Calibrator/Source in
feedback mode to maintain a constant average current. The biasing
line for the array was heavily filtered using both stainless steel
powder filters and commercial filters.\cite{Bladh-RevSciInstr-03}

When a constant bias is applied, charge solitons move through the
array and approach the SET. The space correlation of the Wigner
lattice translates into time correlated tunneling of charges into
the SET at the end of the array. Since the full tunneling charge is
injected into the SET island, the SET acts as a non-linear charge
detector.
Numerical simulations\cite{Likharev-IEEETransMag25-89} show, that
the charging of the SET island occurs quasi-continuously, whereas
the discharge happens abruptly by a tunneling event. Because of the
limited bandwidth of our detector, we can only follow the gradual
charging, but not the much faster tunneling event.

Our limited sensitivity prevents us from discriminating the gradual
charging due to single electrons from that of Cooper pairs, as
either tunneling event gives rise only to a ``click" in the detector
response. However, we can discriminate by frequency; if the current
is carried by electrons, the frequency will be $f_{e}=I/e$, whereas
if it is carried by Cooper pairs it will be $f_{2e}=I/2e$. The power
spectrum of the signal will thus reveal information about the type
of charge carrier.

The upper graph in Fig.\,\ref{fig:Peaks} shows the power spectral
density of the output signal from the mixer for several different
values of magnetic field, when the array is biased with a constant
current of 275\,fA (from top to bottom, $B_{||}$ goes from 100 to
500\,mT in steps of 50\,mT). For a field of 500\,mT the spectrum has
a clear peak at the frequency $f_e$ (the dashed line to the right in
Fig.\,\ref{fig:Peaks}). This peak is due to time correlated
transport of single electrons.\cite{Bylander-Nature-05}
At fields higher than 500\,mT, the $I-V$ characteristic becomes very
steep, and thus the current becomes very sensitive to fluctuations
in the bias and to background charges, as discussed in
Ref.~\onlinecite{Bylander-Nature-05}. Therefore, the
peak in the spectral density is smeared and disappears into the
noise floor.
For decreasing magnetic field, the peak gradually moves to lower
frequencies, and around 200\,mT it appears at $f_{2e}$ (left dashed
line in Fig.\,\ref{fig:Peaks}). At even lower fields,
the peak is smeared and could not be observed below 100\,mT.

\begin{figure}[t]
\centering
 \includegraphics[height=1.2\columnwidth]{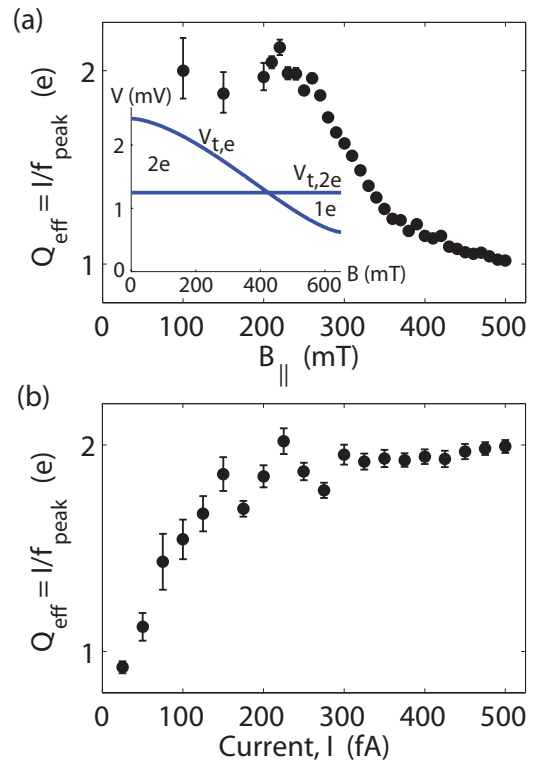}
 \caption{(color online) (a) Effective average charge $Q_\mathrm{eff}=I/f_\mathrm{peak}$
 vs.\@ parallel magnetic field $B_{||}$ for the current $I\!\!=\!200$\,fA
 through the array. For high field, the charge is $1e$, whereas at lower
 field it increases and reaches $2e$ around 200\,mT.
 Inset: Magnetic field dependence of the threshold voltages for
 single electrons ($V_{t,e}$) and Cooper pairs ($V_{t,2e}$),
 see  Eq.~(\ref{eq:thresholds}). In the regions labeled ``$1e$" and
 ``$2e$" only single electrons and single Cooper pairs are allowed,
 respectively. Above both thresholds, both types of charge carriers are allowed.
 (b) Effective charge versus current for $B_{||}=150$\,mT;
 see discussion in the main text.
 The peak frequencies $f_\mathrm{peak}$ and the error bars in
 both plots are obtained from fitting each power spectrum to
 a Lorentzian, see Fig.~2, and $Q_\mathrm{eff}$ is calculated
 using the nominal array current $I$.}
\label{fig:crossover200fA}
\end{figure}

We define an effective charge as the nominal array current divided
by the peak frequency, $Q_\mathrm{eff}=I/f_\mathrm{peak}$, as
obtained from fitting to a Lorentzian and a $1/f^\alpha$ background,
where $\alpha\lesssim1$.
In the intermediate regime, where $1e<Q_\mathrm{eff}<2e$, there is a
mixture of extra single electrons and Cooper pairs in the array.
In Fig.~\ref{fig:crossover200fA}(a), we show how $Q_\mathrm{eff}$
changes as a function of magnetic field for a fixed bias current
$I=200$\,fA. For $B_{||}<250$\,mT, Cooper pair transport dominates;
in the intermediate regime, $250\,\mathrm{mT}<B_{||}<400$\,mT, there
is coexisting $1e$ and $2e$ transport; and for $B_{||}>400$\,mT,
there is predominantly single-electron transport.
In Fig.~\ref{fig:crossover200fA}(b), we show how $Q_\mathrm{eff}$
varies as a function of current for a fixed magnetic field
$B_{||}=150$\,mT. Here, the mixed $1e$ and $2e$ transport occurs in
the region below $I\approx200$\,fA, whereas $Q_\mathrm{eff}=2e$
above this current.

\begin{figure}[t]
\centering
 \includegraphics[height=1.2\columnwidth]{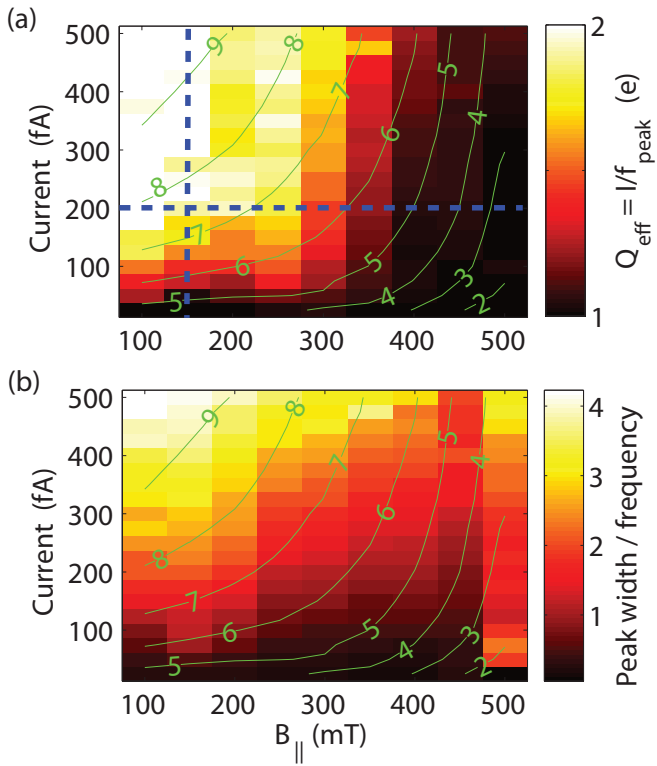}
 \caption{(color online) (a) Effective charge versus magnetic field and current.
 The dashed (blue) lines indicate the cut-outs shown in Fig.~3 for fixed field and current.
 (b) Half width of the peak scaled by peak frequency. The
 solid (green) lines and labels in both plots are contours of constant bias
 voltage in mV across the array.}
\label{fig:Qeff}
\end{figure}

Figure~\ref{fig:Qeff}(a) shows that this magnetic field-induced
crossover occurs only for relatively high current; at low current,
$Q_\mathrm{eff}=1e$ for all magnetic fields. Moreover, this figure
shows that the current-induced crossover occurs only at low magnetic
field, where low current favors electron transport whereas high
current favors Cooper-pairs.
The measured voltages (in mV) across the array are shown as
contours.

In Fig.~\ref{fig:Qeff}(b), we display the normalized width of the
peak (half width divided by frequency) for the same currents and
magnetic fields as in (a). It is clear that the sharpness,
\emph{i.e.}, the degree of correlation between successive tunneling
events, is greater when the transport is dominated by
single-electron tunneling, and for small currents where there are
few solitons inside the array at a given time.


The fact that charge transport with only Cooper pairs is less
correlated than transport with quasiparticles can be qualitatively
explained using energy arguments. The real part of the impedance
seen from a junction inside the array is much smaller than the
quantum resistance $R_Q\!=\!h/4e^2\!\approx\!
6\,\mathrm{k}\Omega$,\cite{Grabert-ZPhysB-91} why energy exchange
with the environment is very ineffective for Cooper pairs.
Therefore, they tunnel in principle only when the charging energy
difference before and after the tunneling event is smaller than
$E_J$. On the contrary, quasiparticles are energetically allowed to
tunnel as soon as the charging energy difference is positive. Thus,
the Cooper pairs are more prone to be trapped inside the array,
degrading the time correlation. Fluctuating background charges, and
also approaching solitons, can change the local bias of a junction
so that the elastic channel opens and the Cooper pair tunnels. In
the regime with mixed $1e$ and $2e$ transport, the peak sharpening
when $Q_\mathrm{eff}$ approaches $1e$ suggests, that the more mobile
$1e$ solitons (quasiparticles) are indeed effective in ``freeing"
the $2e$ solitons (Cooper pairs).

The situation is thus rather complex with a number of things that
affect the transport, including randomly distributed background
charges, non-equilibrium quasiparticles of unknown density, and the
electromagnetic environment. Therefore, a complete quantitative
description of these results is hard to attain. We can, however,
give a number of qualitative arguments to explain the observed
phenomena.

Which type of transport will dominate is largely determined by the
type of carrier that is being injected at the first junction, since
well inside the array, the charges repel each other.
%
The threshold $V_{t,e}$ for injecting a single electron depends on
the magnetic field since the energy $2\Delta$ has to be supplied to
break a Cooper pair. At zero temperature, $V_t$ can be calculated
from electrostatic energy considerations.\cite{footnote} In our
limit $1\ll\Lambda^2\ll N^2$ (and ignoring the effect of random
background charges), we get for electrons and Cooper pairs,
respectively:
\begin{equation} \label{eq:thresholds}
{ V_{t,e}(B)  = \frac{e}{2C_\mathrm{eff}} \left[1+\exp(-1/\Lambda)
\right] \left( 1+\frac{2\Delta(B)}{E_{C'}} \right)
\atop
V_{t,2e} = 2 \frac{e}{2C_\mathrm{eff}} \left[1+\exp(-1/\Lambda)
\right] . }
\end{equation}
Here $C_\mathrm{eff}=\sqrt{C_0^2+4CC_0}=0.23$\,fF is the effective
island capacitance, $E_{C'}=e^2/(2C+C_0+C_\mathrm{eff})=1.7$\,K$k_B$
is the first island's charging energy, and $\Delta(0)/k_B=2.4$\,K 
is the superconducting energy gap of our aluminum thin films at zero
magnetic field and temperature.
The dependence (\ref{eq:thresholds}) is displayed in the inset in
Fig.~\ref{fig:crossover200fA}(a), where we have assumed the
following empirical\cite{Delsing-PRB-94} magnetic field dependence
of the gap: $\Delta(B)/\Delta(0)=(1-(B/B_c)^{1.6})^{1.5}$.
For the given parameters of our sample, $V_{t,e}(0)=2.4$\,mV and
$V_{t,2e}=1.2$\,mV. Background charges will, however, modify these
thresholds. For an array of this size, $V_{t,2e}$ and the part of
$V_{t,e}$ that does not depend on $\Delta$ become approximately
three times larger.\cite{Melsen-PRB-97}
At low field, we therefore expect Cooper pairs to dominate, which is
what we do observe for $I>100$\,fA. At larger fields, $\Delta$ is
suppressed, and therefore also $V_{t,e}$, why at a given field the
single-electron transport becomes significant.

Inside the array, there is also a possibility that a Cooper-pair
soliton centered on one island decays into two single-electron
solitons centered on adjacent islands. This happens when the
difference between the Cooper-pair and single-electron charging
energies is smaller than $2\Delta$. Again disregarding background
charges, the condition for this is
\begin{equation} 
\Delta(B) < \frac{e^2}{2C_\mathrm{eff}} \left[1-\exp(-1/\Lambda)
\right] ,
\end{equation}
which is satisfied for $B_{||}>400$\,mT, meaning that we should
detect pure $1e$ transport for higher fields. This agrees
qualitatively with the data in Fig.~\ref{fig:crossover200fA}(a),
however, we note that thermal fluctuations can break a metastable
Cooper pair at lower fields.

The arguments of the preceding paragraphs explain qualitatively the
magnetic field dependence of the $1e-2e$ crossover at relatively
large currents. Let us now turn to the current dependent crossover
that occurs for $I\lesssim200$\,fA and $B_{||}\lesssim350$\,mT and
consider the different electron and Cooper-pair tunneling rates.
At the temperature of our experiment there should be practically no
thermally excited quasiparticles, but experiments have shown that
there are often non-equilibrium quasiparticles residing in the
leads.\cite{Aumentado-PRL-04}
The threshold voltage for them to enter into the array is lower than
that for Cooper pairs,
$V_{t,e_\mathrm{NE}}=V_{t,e}(\Delta\!=\!0)=V_{t,2e}/2$. This means
that for $V_{t,e_\mathrm{NE}}<V<V_{t,2e}$ only the non-equilibrium
quasiparticles will enter and we should see pure $1e$ transport.
Above $V_{t,2e}$ we expect to see mixed $1e$ and $2e$ transport, and
the different tunneling rates compete.
The quasiparticle tunneling rate is proportional to the number of
non-equilibrium quasiparticles present in the leads. This number is,
in turn, determined through the competition between the process
generating the quasiparticles, their recombination, and the $1e$
current, draining the quasiparticles into the array.
The situation is similar to that of quasiparticle poisoning in the
Cooper pair box, where a similar phenomenon has been
observed\cite{Schneiderman-LT-24} and theoretically
described.\cite{Lutchyn-PRB-05}
This picture qualitatively explains the behavior seen in
Fig.~\ref{fig:crossover200fA}(b). An interesting aspect of this
observation is that it should be possible to extract information
about the density of non equilibrium quasiparticles by making more
elaborate experiments of this kind.

In conclusion, we have demonstrated time correlated tunneling of
both individual electrons and individual Cooper pairs, and
coexistence of the two, in a 1D array of small Josephson junctions.
We have shown that there is a crossover from single-electron
transport to single Cooper-pair transport as a function of both the
external magnetic field and the current through the array. We
describe the transport in terms of different threshold voltages for
injection of charge into the array, and instability of Cooper pairs
inside the array.

We made the sample in the MC2 Nanofabrication Laboratory at Chalmers.
We acknowledge helpful discussions with A. K\"ack, A. Zorin, and 
the members of the Quantum Device Physics Laboratory at MC2. This
work was supported by the Swedish SSF, VR and by the Wallenberg
Foundation.

\bibliographystyle{prb}

\begin{thebibliography}{99}
\bibitem{footnote} We derived Eq.~(\ref{eq:thresholds}) in the same
way as in Ref.~\onlinecite{Bylander-Nature-05}, but here we have
written it in a different form. We note that there is a slight error
in the expression for $V_{t,e}$ in
Ref.~\onlinecite{Bylander-Nature-05}.
\end{thebibliography}

\end{document}